\documentclass[a4paper]{article}
\usepackage{amsmath}

\begin{document}

\title{\textbf{Integrability of the Bakirov system:}\\\textbf{a zero-curvature
representation}}
\author{\textsc{Sergei Sakovich}\bigskip\\{\small Institute of Physics,
National Academy of Sciences, 220072 Minsk, Belarus}\\{\small E-mail:
saks@tut.by}}
\date{}
\maketitle

\begin{abstract}
For the Bakirov system, which is known to possess only one higher-order local
generalized symmetry, we explicitly find a zero-curvature representation
containing an essential parameter.
\end{abstract}

\section{Introduction}

Bakirov \cite{Bak} discovered that the following evidently integrable triangular
system of a linear PDE with a source determined by another linear PDE,
\begin{equation} \label{bak}
u_{t} = u_{xxxx} + v^{2} , \qquad v_{t} = \frac{1}{5} v_{xxxx} ,
\end{equation}
possesses only one higher-order $(x,t)$-independent local
generalized symmetry of order not exceeding 53, namely, a sixth-order one.
Beukers, Sanders, and Wang \cite{BSW} extended the result of Bakirov to
$(x,t)$-independent local generalized symmetries of unlimited order. Bilge
\cite{Bil} found a formal recursion operator for the Bakirov system \eqref{bak}
and showed that the structure of the operator's nonlocal terms prevents the
generation of local higher symmetries from the known sixth-order symmetry.
Sergyeyev \cite{SeO} showed that the existence of such a formal recursion
operator is essentially a consequence of the triangular form of the Bakirov
system. Finally, Sergyeyev \cite{SeJ} proved that the Bakirov system possesses
only one higher-order local generalized symmetry, namely, the sixth-order one
found by Bakirov, even if $(x,t)$-dependent symmetries are taken into account.
Due to these results, the Bakirov system looks quite different from other known
integrable systems which possess infinite algebras of higher symmetries.

In the present paper, we explicitly find a linear spectral problem associated
with the Bakirov system \eqref{bak}, in the form of a zero-curvature
representation containing an essential parameter. Section~\ref{s2} gives
necessary preliminaries. In Section~\ref{s3}, we find for the system \eqref{bak}
a $4 \times 4$ zero-curvature representation containing a parameter, and we prove
in Section~\ref{s4} that this parameter cannot be removed by gauge
transformations. Section~\ref{s5} gives concluding remarks. We believe that the
obtained Lax pair of the Bakirov system will be useful for future studies on the
relation between Lax pairs and higher symmetries of integrable PDEs.

\section{Preliminaries} \label{s2}

A zero-curvature representation (ZCR) of a system of PDEs (see, e.g., \cite{S04}
and references therein) is the compatibility condition
\begin{equation} \label{zcr}
D_{t} X = D_{x} T - \left[ X , T \right]
\end{equation}
of the over-determined linear problem
\begin{equation} \label{lin}
\Psi_{x} = X \Psi , \qquad \Psi_{t} = T \Psi ,
\end{equation}
where $D_{t}$ and $D_{x}$ stand for the total derivatives, $X$ and $T$ are
$n \times n$ matrix functions of independent and dependent variables and
finite-order derivatives of dependent variables, the square brackets denote the
matrix commutator, $\Psi$ is a column of $n$ functions of independent variables,
and the ZCR \eqref{zcr} is satisfied by any solution of the represented system
of PDEs. Two ZCRs are equivalent if they are related by a gauge transformation
\begin{equation} \label{gtr}
\begin{aligned}
X' & = G X G^{-1} + \left( D_{x} G \right) G^{-1} , \\
T' & = G T G^{-1} + \left( D_{t} G \right) G^{-1} , \\
\Psi' & = G \Psi , \qquad \det G \neq 0
\end{aligned}
\end{equation}
of the linear problem \eqref{lin}, where $G$ is a $n \times n$ matrix function
of independent and dependent variables and finite-order derivatives of dependent
variables.

\section{Zero-Curvature Representation} \label{s3}

Our aim is to find a ZCR \eqref{zcr} of the Bakirov system (\ref{bak}). Assuming
for simplicity that $X = X \left( u , v \right)$ and $T = T \left( u , v , u_{x}
, v_{x} , u_{xx} , v_{xx} , u_{xxx} , v_{xxx} \right)$, and using \eqref{bak},
we rewrite \eqref{zcr} in the equivalent form
\begin{equation} \label{ide}
\left( u_{xxxx} + v^{2} \right) \frac{\partial X}{\partial u} + \frac{1}{5}
v_{xxxx} \frac{\partial X}{\partial v} - D_{x} T + \left[ X , T \right] =0.
\end{equation}
Since \eqref{ide} cannot be a system of ODEs restricting solutions of
\eqref{bak}, it must be an identity with respect to $u$ and $v$, and therefore
$u$, $v$ and all derivatives of $u$ and $v$ should be treated as formally
independent quantities in \eqref{ide}. This allows us to solve \eqref{ide} and
obtain the following expressions for the matrices $X$ and $T$:
\begin{equation} \label{fxt}
\begin{aligned}
X & = P u + Q v + R , \\
T & = P u_{xxx} + \frac{1}{5} Q v_{xxx} + \left[ R , P \right] u_{xx} +
\frac{1}{5} \left[ R , Q \right] v_{xx} + \left[ R, \left[ R , P \right]
\right] u_{x} \\
& \qquad + \frac{1}{5} \left[ R , \left[ R , Q \right] \right] v_{x} + \left[ R ,
\left[ R , \left[ R , P \right] \right] \right] u + \frac{1}{5} \left[ R ,
\left[ R , \left[ R , Q \right] \right] \right] v + S ,
\end{aligned}
\end{equation}
where $P$, $Q$, $R$ and $S$ are any $n\times n$ constant matrices satisfying
the following commutator relations:
\begin{equation} \label{rel}
\begin{gathered}
P = - \frac{1}{5} \left[ Q , \left[ R , \left[ R , \left[ R , Q \right] \right]
\right] \right] , \qquad \left[ P , Q \right] = 0 , \\
\left[ P , \left[ Q , R \right] \right] = 0 , \qquad \left[ P , \left[ R , P
\right] \right] = 0 , \qquad \left[ Q , \left[ R , Q \right] \right] = 0 , \\
\left[ \left[ R , P \right] ,\left[ R , Q \right] \right] = 0 , \qquad \left[
P , \left[ R , \left[ R , \left[ R , P \right] \right] \right] \right] = 0 , \\
\left[ \left[ R , P \right] , \left[ R , \left[ R , Q \right] \right] \right]
= 0 , \qquad \left[ P , S \right] + \left[ R , \left[ R , \left[ R , \left[ R ,
P \right] \right] \right] \right] = 0 , \\
\left[ Q , S \right] + \frac{1}{5} \left[ R , \left[ R , \left[ R , \left[ R ,
Q \right] \right] \right] \right] = 0 , \qquad \left[ R , S \right] = 0 .
\end{gathered}
\end{equation}

We have to find a solution of \eqref{rel} which should be nontrivial in the
following sense: $X$ contains both $u$ and $v$, that is, \eqref{zcr} gives
expressions for both equations of \eqref{bak}; $\left[ X , T \right] \neq 0$,
because commutative ZCRs are simply matrices of conservation laws (for this
reason, and without loss of generality, the matrices $P$, $Q$, $R$ and $S$ are
set to be traceless); $X$ contains a parameter (`essential' or `spectral') which
cannot be removed (`gauged out') by gauge transformations \eqref{gtr}. We solve
\eqref{rel}, using the \textit{Mathematica} computer algebra system \cite{Wol},
successively taking $Q$ in all possible Jordan forms, suppressing the excessive
arbitrariness of solutions by transformations \eqref{gtr} with constant $G$,
and increasing the matrix dimension $n$ if necessary. The cases of $2 \times 2$
and $3 \times 3$ matrices contain no nontrivial solutions of \eqref{rel}, while
the $4 \times 4$ case gives us the following:
\begin{gather}
P =
\begin{pmatrix}
0 & 0 & \frac{8}{5} z \left( - 3 + 6 z - 11 z^{2} \right) \alpha^{3} & 0 \\
0 & 0 & 0 & 0 \\
0 & 0 & 0 & 0 \\
0 & 0 & 0 & 0
\end{pmatrix}
, \label{map} \\
Q =
\begin{pmatrix}
0 & 1 & 0 & 0 \\
0 & 0 & 1 & 0 \\
0 & 0 & 0 & 0 \\
0 & 0 & 0 & 0
\end{pmatrix}
, \label{maq} \\
R =
\begin{pmatrix}
\alpha & 0 & 0 & 0 \\
0 & z \alpha & 0 & \alpha \\
0 & 0 & \left( - 1 + 2 z \right) \alpha & 0 \\
0 & \left( - 3 + 6 z - 11 z^{2} \right) \alpha & 0 & - 3 z \alpha
\end{pmatrix}
, \label{mar} \\
S =
\begin{pmatrix}
S_{11} & 0 & 0 & 0 \\
0 & S_{22} & 0 & S_{24} \\
0 & 0 & S_{33} & 0 \\
0 & S_{42} & 0 & S_{44}
\end{pmatrix}
\label{mas}
\end{gather}
with
\begin{equation}
\begin{aligned}
S_{11} & = \frac{8}{5} \left( 2 - 12 z + 21 z^{2} - 18 z^{3} + 3 z^{4} \right)
\alpha^{4} , \\
S_{22} & = \frac{8}{5} \left( 3 - 10 z + 15 z^{2} - 4 z^{4} \right) \alpha^{4}
, \\
S_{24} & = \frac{8}{5} \left( - 1 + 3 z + z^{2} - 3 z^{3} \right) \alpha^{4}
, \\
S_{33} & = \frac{8}{5} \left( - 8 + 28 z - 39 z^{2} + 22 z^{3} - 7 z^{4} \right)
\alpha^{4} , \\
S_{42} & = \frac{8}{5} \left( 3 - 15 z + 26 z^{2} - 18 z^{3} - 29 z^{4} + 33
z^{5} \right) \alpha^{4} , \\
S_{44} & = \frac{8}{5} \left( 3 - 6 z + 3 z^{2} - 4 z^{3} + 8 z^{4} \right)
\alpha^{4} ,
\end{aligned}
\end{equation}
where $\alpha$ is an arbitrary parameter, and $z$ is any of the four roots
\begin{equation} \label{zzz}
\begin{aligned}
z_{1,2} & = \frac{1}{20} \left( 9 + \mathrm{i} \sqrt{39} \pm \sqrt{- 138 - 2
\mathrm{i} \sqrt{39}} \right) , \\
z_{3,4} & = \frac{1}{20} \left( 9 - \mathrm{i} \sqrt{39} \pm \sqrt{- 138 + 2
\mathrm{i} \sqrt{39}} \right)
\end{aligned}
\end{equation}
of the algebraic equation
\begin{equation} \label{eqz}
3 - 12 z + 21 z^{2} - 18 z^{3} + 10 z^{4} = 0 .
\end{equation}

Consequently, a nontrivial ZCR \eqref{zcr} of the Bakirov system \eqref{bak} is
determined by the following $4 \times 4$ matrices $X$ and $T$:
\begin{equation} \label{max}
X =
\begin{pmatrix}
\alpha & v & \frac{8}{5} z \left( - 3 + 6 z - 11 z^{2} \right) \alpha^{3} u & 0
\\
0 & z \alpha & v & \alpha \\
0 & 0 & \left( - 1 + 2 z \right) \alpha & 0 \\
0 & \left( - 3 + 6 z - 11 z^{2} \right) \alpha & 0 & - 3 z \alpha
\end{pmatrix}
\end{equation}
and
\begin{equation} \label{mat}
T =
\begin{pmatrix}
T_{11} & T_{12} & T_{13} & T_{14} \\
0 & T_{22} & T_{23} & T_{24} \\
0 & 0 & T_{33} & 0 \\
0 & T_{42} & T_{43} & T_{44}
\end{pmatrix}
\end{equation}
with
\begin{equation} \label{ttt}
\begin{aligned}
T_{11} & = \frac{8}{5} \left( 2 - 12 z + 21 z^{2} - 18 z^{3} + 3 z^{4} \right)
\alpha^{4} , \\
T_{12} & = \frac{1}{5} \left( - 4 \left( 2 - 3 z + 6 z^{2} + 3 z^{3} \right)
\alpha^{3} v - 2 \left( 1 - 2 z + 5 z^{2} \right) \alpha^{2} v_{x} \right. \\
& \qquad \left. + \left( 1 - z \right) \alpha v_{xx} + v_{xxx} \right) , \\
T_{13} & = \frac{8}{5} z \left( - 3 + 6 z - 11 z^{2} \right) \alpha^{3} \left(
8 \left( 1 - z \right)^{3} \alpha^{3} u + 4 \left( 1 - z \right)^{2} \alpha^{2}
u_{x} \right. \\
& \qquad \left. + 2 \left( 1 - z \right) \alpha u_{xx} + u_{xxx} \right) , \\
T_{14} & = \frac{1}{5} \alpha \left( 4 z \left( - 3 + z \right) \alpha^{2} v - 2
\left( 1 + z \right) \alpha v_{x} - v_{xx} \right) , \\
T_{22} & = \frac{8}{5} \left( 3 - 10 z + 15 z^{2} - 4 z^{4} \right) \alpha^{4}
, \\
T_{23} & = \frac{1}{5} \left( 4 \left( - 2 + 9 z - 18 z^{2} + 19 z^{3} \right)
\alpha^{3} v - 2 \left( 1 - 2 z + 5 z^{2} \right) \alpha^{2} v_{x} \right. \\
& \qquad \left. + \left( 1 - z \right) \alpha v_{xx} + v_{xxx} \right) , \\
T_{24} & = \frac{8}{5} \left( - 1 + 3 z + z^{2} - 3 z^{3} \right) \alpha^{4} , \\
T_{33} & = \frac{8}{5} \left( - 8 + 28 z - 39 z^{2} + 22 z^{3} - 7 z^{4} \right)
\alpha^{4} , \\
T_{42} & = \frac{8}{5} \left( 3 - 15 z + 26 z^{2} - 18 z^{3} - 29 z^{4} + 33 z^{5}
\right) \alpha^{4} , \\
T_{43} & = \frac{1}{5} \left( - 3 + 6 z - 11 z^{2} \right) \alpha \left( 4 z
\left( - 3 + 5 z \right) \alpha^{2} v \right. \\
& \qquad \left. + \left( 2 - 6 z \right) \alpha v_{x} + v_{xx} \right) , \\
T_{44} & = \frac{8}{5} \left( 3 - 6 z + 3 z^{2} - 4 z^{3} + 8 z^{4} \right)
\alpha^{4} .
\end{aligned}
\end{equation}
Let us remind that, in \eqref{max} and \eqref{ttt}, $z$ is any of the four roots
\eqref{zzz} of the equation \eqref{eqz}, and $\alpha$ is an arbitrary parameter.

\section{Essential Parameter} \label{s4}

Now, we have to prove that $\alpha$ is an essential parameter, that is, that
$\alpha$ cannot be removed from the obtained ZCR by a gauge transformation
\eqref{gtr}. We do this, using the method of gauge-invariant description of
ZCRs of evolution equations \cite{S04,S95} (see also the independent work
\cite{M97}, based on the very general and abstract study of ZCRs \cite{M92}).
Since the matrix $X$ \eqref{max} does not contain derivatives of $u$ and $v$,
the two characteristic matrices of the obtained ZCR are simply $C_{u} =
\partial X / \partial u = P$ and $C_{v} = \partial X / \partial v = Q$. We
take one of them, $C_{u} = P$ \eqref{map}, introduce the operator $\nabla_{x}$,
defined as $\nabla_{x} M = D_{x} M - \left[ X , M \right]$ for any $4 \times 4$
matrix function $M$, compute $\nabla_{x} C_{u}$, and find that
\begin{equation} \label{clo}
\nabla_{x} C_{u} + 2 \left( 1 - z \right) \alpha C_{u} = 0 .
\end{equation}
In the terminology of \cite{S04,S95}, the relation \eqref{clo} is one of the two
closure equations of the cyclic basis. The scalar coefficient $2 \left( 1 - z
\right) \alpha$ in \eqref{clo} is an invariant with respect to the gauge
transformations \eqref{gtr}, because the matrices $C_{u}$ and $\nabla_{x} C_{u}$
are transformed as $C_{u}^{\prime} = G C_{u} G^{-1}$ and $\nabla_{x}^{\prime}
C_{u}^{\prime} = G \left( \nabla_{x} C_{u}\right) G^{-1}$ (see \cite{S95} and
\cite{M97}). The explicit dependence of the invariant $2 \left( 1 - z \right)
\alpha$ on the parameter $\alpha$ shows that this parameter cannot be `gauged
out' from the matrix $X$ \eqref{max}.

\section{Conclusion} \label{s5}

We believe that the ZCR of the Bakirov system, found in this paper, can be used
in future studies of the relation between Lax pairs, recursion operators,
generalized symmetries, and conservation laws. The following problems arise from
the obtained result. Is it possible to derive a recursion operator for the
Bakirov system from the obtained ZCR, for example, through the cyclic basis technique
\cite{S04}? If yes, is that recursion operator different from the formal
recursion operator found in \cite{Bil}? And why does not the obtained ZCR
generate an infinite sequence of nontrivial local conservation laws for the
Bakirov system, for example, through the standard techniques described in \cite{STS}?

\section*{Acknowledgments}
The author is grateful to Artur Sergyeyev for useful information about symmetries
of the Ibragimov--Shabat and Bakirov systems, and to the anonymous reviewer for
valuable suggestions.

\end{document}